\documentclass[twocolumn,showpacs,preprintnumbers,prl]{revtex4}

\input{tcilatex}

\begin{document}

\textbf{\textit{Murzin and Jansen Reply:}} In our Letter \cite{MJC} we
describe the observation of oscillatory variations of the Hall ($G_{xy}$)
conductance in view of topological scaling effects giving rise to the
quantum Hall effect. Such oscillations are experimentally observed in
disordered GaAs layers in the extreme quantum limit of applied magnetic
field. They occur in a field range without oscillations of the density of
states due to Landau quantization and are, therefore, totally different from
the Shubnikov-de Haas oscillations. In a more detailed analysis of the
observed oscillations, we have made a quantitative comparison with
conceptually different theories based on a phenomenological approach \cite%
{Dolan} and on a microscopic description \cite{PB95} of the scaling problem.
In a comment \cite{comment}, Pruisken and Burmistrov question the
microscopic justification of the phenomenological approach and the
applicability of the microscopic theory to our experimental data.

Certainly, the phenomenological approach \cite{Dolan} is hypothetical and
does not have microscopic justification. However, this does not mean that
the results of this approach cannot be valid and experimental comparison
will be of interest. The phenomenological approach \cite{Dolan} has been
developed for macroscopic samples at finite temperatures. Our derived
results for the Hall-conductance oscillation amplitudes from the general
expressions of the unified scaling theory \cite{Dolan} hold in the limit $%
\exp (-2\pi G_{xx})\ll 1$ for the case of small amplitudes of the
oscillations of $G_{xx}$ and $G_{xy}$ with respect to both $1$ (in units $%
e^{2}/h$)\ and $G_{xx}^{sm}(T)-G_{xx}^{0}$, where $G_{xx}^{sm}(T)$ is the
smooth part of diagonal conductance and $G_{xx}^{0}$ is the bare diagonal
conductance. These conditions are valid in our experiments. The oscillations
of $G_{xy}$ are quantitatively well described by this theoretical approach.

In our discussion of the "dilute instanton gas" approximation (DIGA) \cite%
{MJC} we didn't consider the difference between the macroscopic conductances
measured at finite temperature and the ensemble averaged conductances at $%
T=0 $ used in this microscopic approach \cite{PB95}. The very large
difference (up to 10 times) in the oscillation amplitudes between experiment
and theory (extracted from Ref.\cite{PB95} using the ensemble averaged
conductance at $T=0$) has led us to the conclusion that the DIGA results
differ quantitatively from the experimental data. More recently \cite{PB05},
Pruisken and Burmistrov obtained a roughly 5 times smaller topological
oscillatory term (presented in the comment) than that derived in Ref.\cite%
{PB95} which would be in line with our conclusion concerning the smallness
of the measured amplitude oscillations with respect to the result \cite{PB95}
as known at the moment of publication. Since an explicit computation of the
function $f_{H}(g_{0})$ has not been done \cite{comment}, we cannot give a
quantitative comparison of the oscillation amplitudes with our experimental
data considering the DIGA approach for the measured conductance.

Note, that in our experiments $G_{xx}^{sm}$ is a logarithmic function of the
temperature with the same coefficient close to $1/\pi $ in front of the
logarithm for all four samples in the 2D regime ($T<1$ K). Such behavior is
in line with the theory of the quantum corrections due to dominant
electron-electron interactions, although for $G_{xx}^{sm}\sim 1$ one would
expect higher order contributions beyond the logarithmic temperature
dependence. Using Eq.(2) of the $\beta $-function \cite{comment} for the
range of values of our experimental data, the smooth part of the ensemble
averaged conductance $\sigma _{0}$ is essentially outside the quantum
corrections limit with logarithmic temperature dependence because, for $%
\sigma _{0}\sim 1$, the term $\propto 1/\sigma _{0}$ becomes comparable with
the first constant term in this equation assigning the logarithmic behavior.

\bigskip

S.S. Murzin

\textit{Institute of Solid State Physics RAS, 142432, Chernogolovka, Moscow
District, Russia}

\medskip

A.G.M. Jansen

\textit{Service de Physique Statistique, Magn\'{e}tisme, et Supraconductivit%
\'{e}, D\'{e}partement de Recherche Fondamentale sur la Mati\`{e}re Condens%
\'{e}e, CEA-Grenoble, 38054 Grenoble Cedex 9, France}

\end{document}